\documentclass[final, 5p, twocolumn, times]{elsarticle}

\usepackage{lineno,hyperref}
\usepackage{subcaption}
\usepackage{float}
\usepackage[utf8]{inputenc}
\modulolinenumbers[1]
\journal{Nuclear Instruments and Methods in Physics Research Section A}
\begin{document}

\begin{frontmatter}

\title{Performance analysis of the prototype THz-driven electron gun for the AXSIS project.}
\author[m1]{G. Vashchenko\corref{mycorrespondingauthor}}
\cortext[mycorrespondingauthor]{Corresponding author}
\ead{grygorii.vashchenko@desy.de}
\author[m1]{R. Assmann}
\author[m1]{U. Dorda}
\author[m1,m2]{M. Fakhari}
\author[m1]{A. Fallahi}
\author[m1]{K. Galaydych}
\author[m1,m2,m3]{F. Kaertner}
\author[m1]{B. Marchetti}
\author[m1]{N. Matlis}
\author[m1]{T. Vinatier}
\author[m1]{W. Qiao}
\author[m1,m2]{C. Zhou}
\address[m1]{Deutsches Elektronen Synchrotron (DESY), Notkestrasse 85, 22607 Hamburg, Germany}
\address[m2]{Department of Physics, University of Hamburg, Luruper Chaussee 149, 22761 Hamburg, Germany}
\address[m3]{The Hamburg Centre for Ultrafast Imaging, Luruper Chaussee 149, 22761 Hamburg, Germany}

\begin{abstract}
The AXSIS project (Attosecond X-ray Science: Imaging and Spectroscopy) aims to develop a THz-driven compact X-ray source for applications e.g. in chemistry and biology by using ultrafast coherent diffraction imaging and spectroscopy. The key components of AXSIS are the THz-driven electron gun and THz-driven dielectric loaded linear accelerator as well as an inverse Compton scattering scheme for the X-rays production.
\par
This paper is focused on the prototype of the THz-driven electron gun which is capable of accelerating electrons up to tens of keV. Such a gun was manufactured and tested at the test-stand at DESY. Due to variations in gun fabrication and generation of THz-fields the gun is not exactly operated at design parameters. Extended simulations have been performed to understand the experimentally observed performance of the gun. A detailed comparison between simulations and experimental measurements is presented in this paper.
\end{abstract}


\end{frontmatter}

\section{Introduction}
The AXSIS project aims to develop a compact THz-driven X-ray source. An electron beam will be produced in a THz-driven electron gun, further accelerated up to $\mathrm{20 \, MeV}$ in a THz-driven dielectric-loaded waveguide and used in an Inverse Compton Scattering system to obtain ultra-short highly coherent X-ray pulses~\cite{AXSIS}.
\par
A prototype of a THz-driven photoelectron gun, the so-called "Horn gun", was developed and tested at DESY. This gun was originally designed for the use with Gaussian-modulated single cycle THz pulses with a central frequency of $\mathrm{450 \, GHz}$, $\mathrm{17 \, \mu J}$ pulse energy and pulse duration of $\mathrm{2.5 \, ps}$. A transverse Gaussian distribution with the rms beam waist radius of $\mathrm{0.5 \, mm}$ was used for the design. A gun sketch with a truncated big horn can be found in the "Simulations" section, more details on the gun design can be found in~\cite{Fallahi,Vashchenko}. With these parameters and by using a UV laser as a photoelectron source this gun was designed to deliver electrons with about $\mathrm{25 \, keV}$ energy.
\par
Due to various reasons such as the present limit of the available THz energy of about $\mathrm{4 \, \mu J}$, the difficulty in focusing the THz beam to the transverse size smaller than $\mathrm{1 \, mm}$ and others, the gun could not be operated at the design parameters. Additionally, due to technical reasons caused by tiny dimensions of the gun, some details of the gun geometry had to be changed during the production. Thus, a reduction of the gun performance was expected.

\section{Experimental results}
\begin{figure*}[!htb]
    \centering
    \includegraphics[width=0.75\textwidth, height=9.96cm]{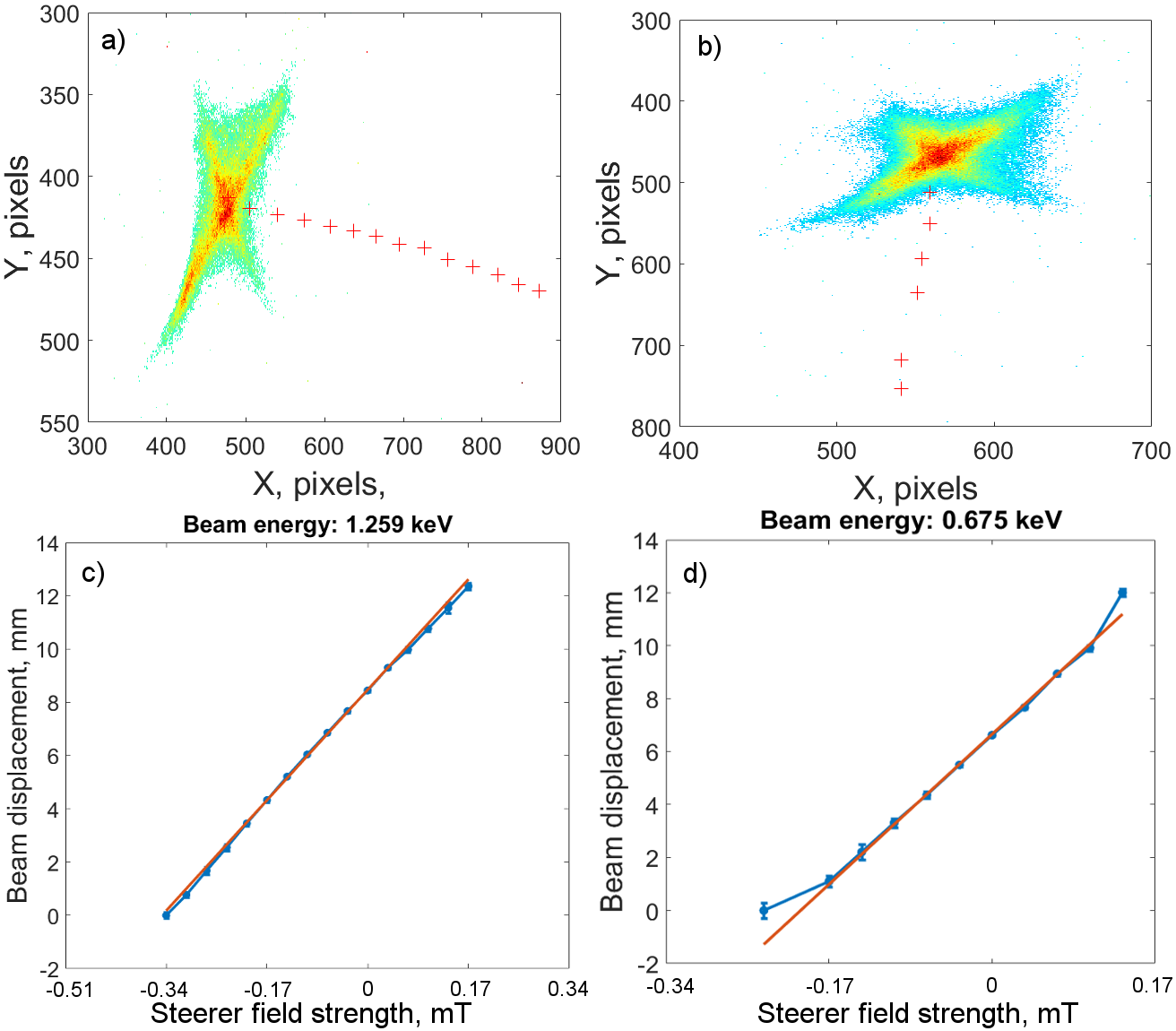}
    \caption{Electron beam energy measurement. Crosses on top graphs represent the beam center movement on the screen vs. dipole current while using the horizontal or vertical deflection. Bottom figures show the analysis of the data from top figures: lines with markers represent experimental data, solid lines - linear fit of the experimental data.}
    \label{Fig:Energy}
\end{figure*}
A set of diagnostic devices was procured and used at the experimental test-stand located at Center for Free-Electron Laser Science (CFEL), DESY. At the first stage, an electron multiplier installed directly at the gun chamber exit was used to detect electrons. With this device we performed a bunch charge scan of the time-delay between the UV-pulse and the THz driver as shown in~\cite{Vashchenko}.
\par
As a next step, a microchannel plate detector (MCP) was used to record the electron beam transverse profile. Finally, the diagnostic line was upgraded with electrostatic plates, a steering dipole and an air coil solenoid. Using this diagnostic line the electron beam was characterized in terms of charge, transverse distribution and energy as shown in~\cite{Vashchenko}. For the energy measurement the PCB steerer~\cite{Flottmann}, solenoid and MCP were used. The current of the dual plane steerer was scanned in the horizontal and vertical planes and the beam distributions were recorded using the MCP. The solenoid was used to focus the electron beam onto a scintillating screen, which is coupled to the MCP using fiber optics. The most recent analysis of the experimental data is presented in Fig.~\ref{Fig:Energy}.
\par
Red crosses on Figures~\ref{Fig:Energy}~a) and b) show the beam movement on the screen while varying the current of the steerer in the horizontal and vertical planes, respectively. The beam distributions are shown only for the final value of the steerer current. Figures~\ref{Fig:Energy}c) and d) show the beam displacement on the screen as a function of steerer current and are obtained from the results of Figures~\ref{Fig:Energy}.a) and b) applying the camera calibration of $\mathrm{26.5 \, \mu m / pixel}$. The energy calibration of the steerer was obtained by measuring the displacement of an electron beam with predefined energy on the observation screen, which was performed at the Photo-injector Test Facility at Zeuthen, DESY.
\par
For the energy measurements using the horizontal deflection of the beam, the obtained beam displacement vs. the steerer current is, as expected, linear and results in the beam energy of about $\mathrm{1.3 \, keV}$. During the measurements in the vertical plane the input THz power was not perfectly stable and the linear fit is not as good as for the horizontal plane, the estimated beam energy is about $\mathrm{0.7 \, keV}$. These beam energy values are significantly lower than what would be expected due to variation of the experimental parameters and gun geometry changes as compared to the design case. Therefore an extended set of simulations for the manufactured gun was launched to further understand the electron beam dynamics under the experimental conditions.

\section{Simulations}
\begin{figure*}[!htb]
    \centering
    \includegraphics[width=0.9\textwidth]{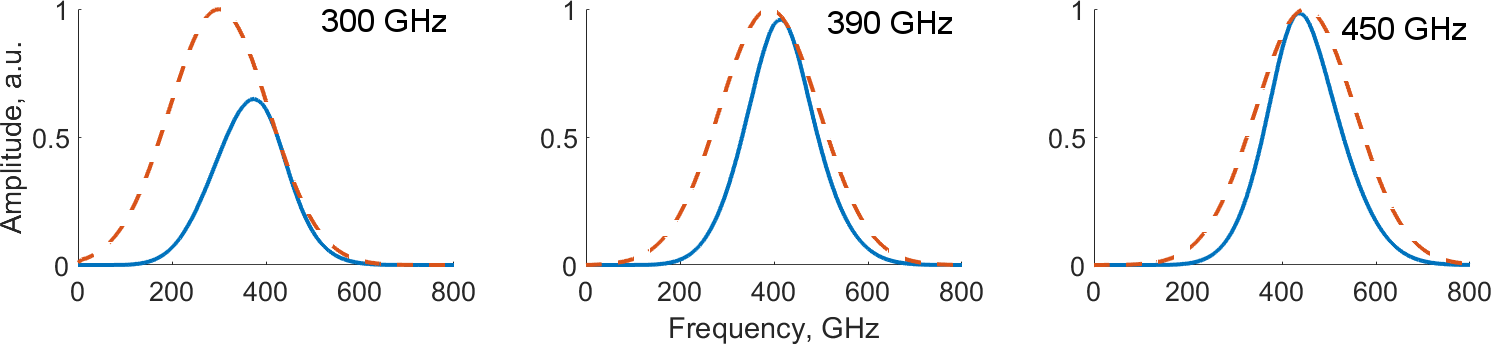}
    \caption{Gun response for central frequencies of the initial THz pulse (from left to right) of 300, 390 and $\mathrm{450 \, GHz}$. Dashed lines represent spectra of the initial THz pulse, solid lines represent the gun response at the cathode.}
    \label{Fig:FFT}
\end{figure*}

Various simulations of the gun were performed using the LEIST code which is a three-dimensional hybrid technique based on discontinuous Galerkin time domain and particle in cell methods~\cite{LEIST}.
\paragraph{Frequency domain} As mentioned above, the gun was originally designed to operate at the central frequency of $\mathrm{450 \, GHz}$. In the experimental case it was only possible to operate the gun at the central frequency of either $\mathrm{300}$ or $\mathrm{390 \, GHz}$. To understand the impact of the different frequencies on the gun performance field simulations were conducted. Gun response at the cathode for above mentioned frequencies was compared to the initial THz pulse spectrum as shown in Fig.~\ref{Fig:FFT}.
For this simulation a THz pulse with a Gaussian-modulated longitudinal distribution, $\mathrm{2.5 \, ps}$ duration (wide band) and rms beam waist radius of $\mathrm{0.5 \, mm}$ was used. The analysis was performed only on the longitudinal (accelerating) component of the electric field at the center of the photocathode.
\par
As we can see from the plots, the most efficient response of the gun is obtained for frequencies between 390 and $\mathrm{450 \, GHz}$. At $\mathrm{300 \, GHz}$ the gun performance is limited as the major fraction of the pulse is situated below the cut-off frequency of the accelerating channel and is therefore attenuated. For central frequencies of 390 and $\mathrm{450 \, GHz}$ the attenuation below $\mathrm{\sim 400 \, GHz}$ is low but a small reduction of the pulse amplitude for frequencies above $\mathrm{\sim 400 \, GHz}$ is observed due to field leakage through the output slit.

\paragraph{Transmission properties} As explained in~\cite{Vashchenko}, the metallic wall at the end of the smaller horn serves as a reflector for the THz pulse and further enhances the electric field at the cathode, consequently leading to higher electron beam energies. This wall, hereinafter "mirror", was produced separately from the gun body and installed on a micromover. This added the possibility of fine tuning between forward and reflected THz pulses by changing the position of the mirror. During the experiment no notable difference in the electron beam properties was observed for different positions of the mirror in a wide range (more than a wavelength). In order to understand this effect, a set of simulations on the gun transmission was launched.
\par
\begin{figure}[!htb]
    \centering
    \includegraphics[width=0.45\textwidth]{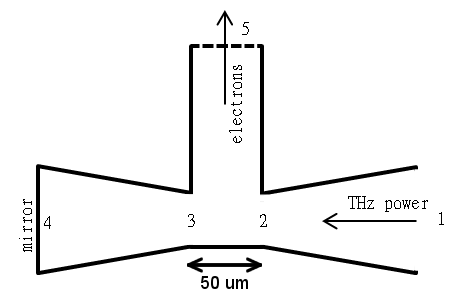}
    \caption{Schematic view of the gun in the vicinity of the accelerating channel. The numbers indicate the positions of the field monitors used in the simulations for recording the electromagnetic fields: 1 - big horn entrance, 2 - accelerating channel entrance, 3 - accelerating channel exit, 4 - mirror, 5 - gun output.}
    \label{Fig:Poynting}
\end{figure}

To perform the analysis of the THz energy transmission, we recorded the energy of the pulse at different points in the gun. For this purpose, several monitors of electric and magnetic fields were used in the simulations. The locations of the monitors are shown in Fig.~\ref{Fig:Poynting}. The energy $W$ was then calculated using the Poynting vector and integrating over the plane and time: $W = \int \int [\vec{E} \times \vec{H}] d\vec{S}dt$.
Incoming THz pulses with $\mathrm{5 \, \mu J}$ energy, $\mathrm{0.5 \, mm}$ rms radius at the waist, $\mathrm{2.5 \, ps}$ pulse duration and central frequencies of $\mathrm{300 \, GHz}$ and $\mathrm{450 \, GHz}$ were used. The results of the simulation are presented in Table~\ref{Table:Poynting}.
\begin{table}
\centering
\begin{tabular}{l | c | c}
    \hline
    Position & Energy@450GHz, $\mathrm{\mu J}$ & Energy@300GHz, $\mathrm{\mu J}$  \\
    \hline
    1 & 4.86 & 4.54\\
    2 & 1.7 & 1.13\\
    3 & 1.1 & 0.49\\
    4 & 1.1 & 0.48\\
    5 & 0.52 & 0.09\\
    \hline
\end{tabular}
\caption{Results of the transmission simulations.}
\label{Table:Poynting}
\end{table}
\par
As we can see for the aforementioned THz pulse parameters a fraction of the energy is lost already at the input as the transverse size of the THz pulse is larger than the input aperture of the gun. Also we can notice that the amount of energy transferred from point 2 to point 3 is significantly higher for the THz frequency of $\mathrm{450 \, GHz}$. Moreover, the total energy transferred from point 2 onward is close to $\mathrm{100 \, \%}$, with a fraction of the energy leaking through the output slit. While leaking through the slit the electric field transforms so that there is a transverse component which causes the electron beam kick. Simulation shows that for the $\mathrm{450 \, GHz}$ case the kick angle can reach $\mathrm{10 \, deg}$ and leads to charge losses on the slit walls. In case of $\mathrm{300 \, GHz}$, we observe only a relatively low amount of energy transferred from point 2 to point 3. Therefore no significant enhancement of the accelerating field with respect to the position of the mirror can be expected. These agree with the results of the frequency domain simulations and matches the experimental observations.

\begin{figure}[!htb]
    \centering
    \includegraphics[width=0.45\textwidth]{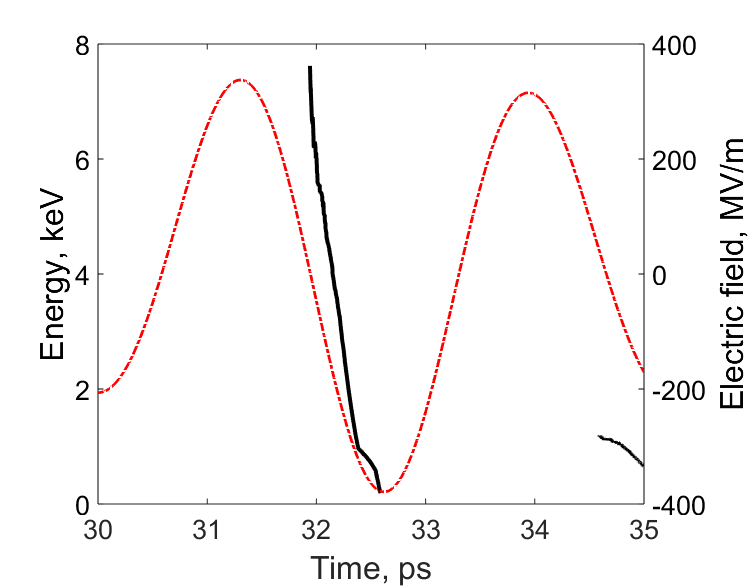}
    \caption{Example of the single simulation for the THz pulse with an energy of $\mathrm{13 \, \mu J}$ and central frequency of $\mathrm{300 \, GHz}$.
     As there is no space-charge forces included in the simulation such a plot gives an information about the maximum energy that can be gained by the electron and the proper injection time of the UV laser with respect to the THz pulse. Solid line represents final electron beam energy, dashed lines shows the electric field profile at the cathode.}
    \label{Fig:Example}
\end{figure}

\paragraph{Single particle simulations for different gradients} In order to better understand the performance of the gun, a set of simulations for various THz pulse energies was launched. Simulations were conducted for different central frequencies, with and without the mirror. For these simulations 50 thousand macroparticles were uniformly distributed in time to cover the entire range when there is an electric field in the accelerating channel (the time profile of the electric field can be found in~\cite{Vashchenko}). Spatial distribution was chosen to be Gaussian with the rms radius of $\mathrm{5 \, \mu m}$. The space-charge forces were switched off. As a result, the final electron energies for a fixed THz energy and frequency were obtained. The result of such simulation at the vicinity of the lowest electric field peak, which must yield the highest acceleration, is depicted in Fig.~\ref{Fig:Example}. The THz pulse energy of $\mathrm{13 \, \mu J}$ and the central frequency of $\mathrm{300 \, GHz}$ were used in this simulation.
\begin{figure}[!htb]
    \centering
    \includegraphics[width=0.45\textwidth]{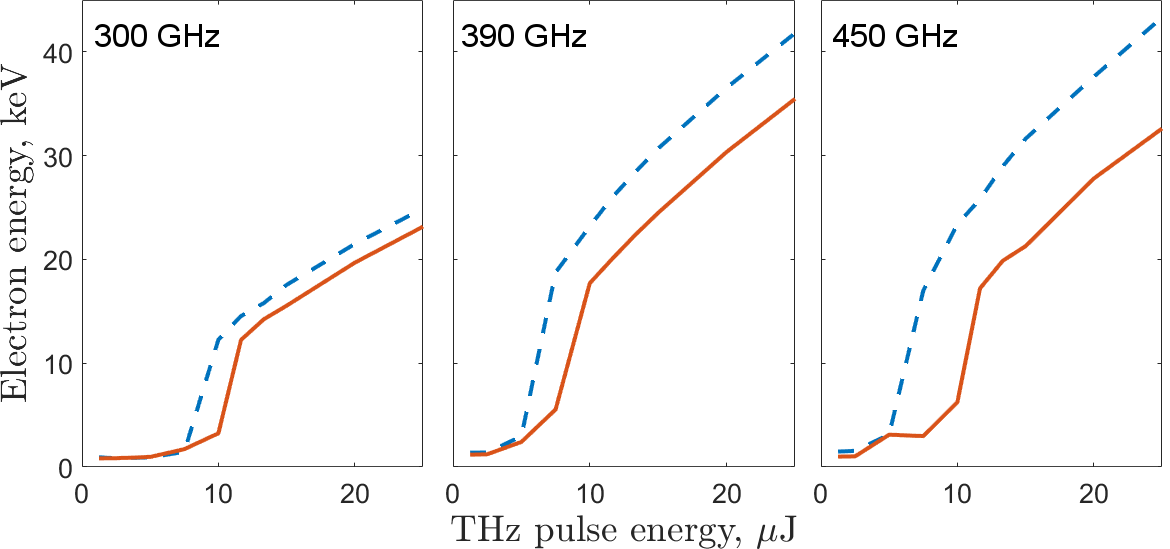}
    \caption{Summary of the simulations for the maximum electron kinetic energy vs. input THz energy for different frequencies, from left to right: 300, 390 and 450 GHz. Dashed lines: simulation with the mirror, solid lines: simulation without the mirror.}
    \label{Fig:Gradient}
\end{figure}

From such a plot we can obtain the maximum electron energy gain for the fixed THz pulse energy and frequency, including or excluding the mirror. By running those simulations for central frequencies of 300, 390 and $\mathrm{450 \, GHz}$ for various THz pulse energies, the following results were obtained and are presented in Fig.~\ref{Fig:Gradient}.
\par
As we can see for the central frequency of $\mathrm{300 \, GHz}$ there is only a small difference in the energy gain with and without the mirror which matches the expectations from the transmission simulations. As compared to $\mathrm{300 \, GHz}$, for the central frequency of $\mathrm{450 \, GHz}$ a significant energy gain is seen while operating with the mirror. Also we can see that for all cases there is a threshold on the input THz energy below which we can not obtain an efficient acceleration (e.g. sharp rise of the final electron energy in the left plot of Fig.~\ref{Fig:Gradient} at about $\mathrm{10 \, ps}$). This point needs a detailed explanation. Firstly, we have to take into account that the accelerating field in the gun is produced by single cycle THz pulse which disperses during the propagation in the gun towards the accelerating channel. Due to this effect the electric field in the accelerating channel consists of several peaks as shown in~\cite{Vashchenko}. In the general case, the most efficient acceleration is obtained when the particle is launched on-crest with the largest negative peak of the field. Once the field amplitude is high enough, the electron experiences acceleration and leaves the gun within the negative half-cycle of the field. This is the region of the THz pulse energies above $\mathrm{\sim 10 \, \mu J}$ in Fig.~\ref{Fig:Gradient}.
\par
For THz pulse energies of about $\mathrm{\sim 6 \, \mu J}$ and below, the electron cannot escape the gun within the negative half-cycle of the acceleration field and slips over the remaining THz pulse, experiencing continuous acceleration and deceleration in several upcoming THz pulse cycles. As the amplitude of the field decreases fast from cycle to cycle, an integrated over the rest of the THz pulse electric field is small and almost independent on the maximum pulse amplitude. Therefore, the final energy gain weakly depends on the THz pulse energy and is close to about $\mathrm{1 \, keV}$ for the simulated parameters, independently from the central frequency of the THz pulse. Also there is no effect of the mirror on the electron energy gain can be expected in this case. This agrees with the results of the corresponding transmission simulations and the experimental observations. Also we can see that gun operation with the mirror reduces requirements on the amount of THz energy which is needed to obtain efficient acceleration.

\section{Summary}
A THz gun driven by a single cycle THz pulse was successfully put into operation at the gun test-stand. Due to various reasons the experimental conditions did not match the design case and low performance of the gun was observed experimentally. Namely, the measured electron beam energy was about $\mathrm{1.3 \, keV}$ which is significantly lower than what was originally expected, even taking into account worse than design case experimental parameters. Therefore extended simulations were conducted in order to understand the observed gun performance. Results of these simulations gave a deep understanding of the electron beam dynamics in the gun and showed good agreement with experimental data.
\par
A new laser system capable to deliver significantly more THz energy is currently under development. Also a new type of the gun (single layer butterfly gun, see~\cite{Fallahi} for conceptual details) is in preparation. With this upgraded setup we are looking forward to achieving significantly higher electron energy gains as compared to the Horn gun.

\section{Acknowledgement}
The research leading to these results has received funding from the European Research Council under the European Union’s Seventh Framework Programme (FP/2007-2013) / ERC Grant Agreement N. 609920.

\section{References}
\bibliography{mybibfile}

\end{document}